\documentclass[aps,pra,twocolumn,superscriptaddress,amsmath,amssymb,showpacs]{revtex4}
\usepackage{graphicx}
\begin{document}

\title{On the Saturation of the Nonlinear Refractive Index in Atomic Gases}\
\author{C.~K\"ohler}\affiliation{CEA-DAM, DIF, F-91297 Arpajon, France}
\author{Roland~Guichard}\affiliation{CNRS, UMR 7614, LCPMR, 75231 Paris Cedex 05, France}
\author{Emmanuel~Lorin}\affiliation{Carleton University, Ottawa (Ontario) K1S 5B6, Canada}
\author{Szczepan~Chelkowski}\affiliation{D\'epartement de chimie, Universit\'e de Sherbrooke, Sherbrooke (Qu\'ebec) J1K 2R1, Canada}
\author{Andr\'e~D.~Bandrauk}\affiliation{D\'epartement de chimie, Universit\'e de Sherbrooke, Sherbrooke (Qu\'ebec) J1K 2R1, Canada}
\author{Luc~Berg\'e}\affiliation{CEA-DAM, DIF, F-91297 Arpajon, France}
\author{Stefan~Skupin}\affiliation{Max Planck Institute for the Physics of Complex Systems, 01187 Dresden, Germany}
\affiliation{Friedrich Schiller University, Institute of Condensed Matter Theory and Optics, 07743 Jena, Germany}

\begin{abstract}
Motivated by the ongoing controversy on the origin of the nonlinear index saturation and subsequent intensity clamping in femtosecond filaments, we study the atomic nonlinear polarization induced by a high-intensity and ultrashort laser pulse in hydrogen by numerically solving the time dependent Schr\"odinger equation.  Special emphasis is given to the efficient modeling of the nonlinear polarization at central laser frequency corresponding to 800~nm wavelength. Here, the recently proposed model of the Higher-Order Kerr Effect (HOKE) and two versions of the Standard model for femtosecond filamentation, including either a multi-photon or tunnel ionization rate, are compared. We find that around the clamping intensity the instantaneous HOKE model does not reproduce the temporal structure of the nonlinear response obtained from the quantum mechanical results. In contrast, the non-instantaneous charge contributions included in the Standard models ensure a reasonable quantitative agreement.  Therefore, the physical origin for the observed saturation of the overall electron response is confirmed to mainly result from contributions of free or nearly free electrons.
\end{abstract}
\maketitle

\section{Introduction}

Accurate modeling of high intensity laser pulse propagation in a medium requires the accurate knowledge of its optical response. For laser intensities well below the atomic ionization threshold this response is given by the induced polarization involving bound electrons, usually modeled by a Taylor expansion in terms of the driving electric field. Restriction of the expansion to the third order nonlinear term for isotropic media leads to the well known intensity-dependent Kerr refractive index change \cite{Shen:PNO:84}. This simplest approach for the polarization already provides all key ingredients to describe fundamental nonlinear phenomena like modulational instability, self-focusing and self-phase modulation \cite{Boyd:NO:1992}, which manifest in soliton formation, pulse compression or self-similar collapse. The latter effect, however, demonstrates the inherent incompleteness of this simplest approach as it predicts an unphysical blow-up of the beam intensity, which is intimately linked to the onset of laser filamentation. 

A theoretical description of laser filamentation therefore requires an accurate modeling of the saturation of the optical nonlinearity, i.e., the induced variation of the nonlinear refractive index with respect to the laser intensity. Originally achieved by accounting for photo-ionization and subsequent interaction of the laser pulse with a free electron plasma \cite{Berge:RPP:70:1633,Berge:DCDS:23:1099,Couairon:pop:7:193,Couairon:PR:441:47,Chin:LP:22:1,Mlejnek:OL:23:382,Brodeur:OL:22:304}, this so-called Standard model for nonlinear refractive index saturation was recently challenged by the measurement of higher order Kerr coefficients \cite{Loriot:OE:17:13429,Loriot:OE:18:3011}. Being negatively valued, these coefficients could give rise to an alternative explanation for the saturation of the optical nonlinearity. This strikingly simple approach, based on higher order contributions in the Taylor expansion of the polarization in terms of the driving electric field, is called the Higher-Order Kerr Effect (HOKE) model. Besides the apparent differences in the model equations, there is a fundamental physical difference between HOKE and Standard model. While in the latter model a free electron plasma is responsible for the arrest of the self-focusing, HOKE terms originate exclusively from the response of bound electrons. Thus, the HOKE model questions the very nature of the previous understanding of femtosecond filamentation as a dynamical balance between optical Kerr effect and self-generated electron plasma \cite{Bejot:PRL:104:103903,Bejot:arxiv:2012,Kolesik:OL:35:2550,Wang:OE:18:24301,Bree:PRL:106:183902,Wahlstrand:OL:36:3822,Nurhuda:NJP:10:053006,Teleki:PRA:82:065801,Volkova:JETPL:94:519}. From the experimental side, numerous attempts have been undertaken to decide in favor for either bound or ionized electron contributions \cite{Polynkin:PRL:106:153902,Chen:PRL:105:215005,Wahlstrand:PRL:107:103901,Wahlstrand:JQE:48:760,Kolesik:OL:35:3685,Kosareva:OL:36:1035,Bejot:PRL:106:243902,Bejot:OL:36:828} without common agreement. Recent progress was achieved by direct numerical solution of the time dependent Schr\"odinger equation (TDSE), which showed the excitation of atoms into highly excited Rydberg and Kramers-Henneberger states. Although bound, these states lie at the frontier above which excited electrons may be fully freed, and they behave very similarly to states in the continuum \cite{Volkova:JETPL:94:519,Nurhuda:NJP:10:053006,Bejot:arxiv:2012}. 

Here, we propose a direct confrontation of both HOKE and Standard models with rigorous quantum mechanical calculations. To this end, we investigate the saturation of the nonlinear polarization by numerically solving the TDSE for an atomic hydrogen model, for the sake of simplicity first in one dimension (1D), and later in full three dimensions (3D) for confirmation. It turns out  that a consistent definition of the HOKE coefficients is difficult, because the validity of the Taylor expansion of the nonlinear response in terms of the electric field breaks down at intensities well below the clamping one. Moreover, the instantaneous character of the HOKE model contradicts the quantum mechanical results. In contrast, we achieve reasonable agreement when applying the Standard model with multi-photon or tunnel ionization rate. Although the exact definition of bound and ionized electrons remains ambiguous for atoms in strong laser fields, the saturating contribution to the nonlinear refractive index is still well described by a simple Drude plasma model assuming continuum electrons as encountered once the pulse has passed the interaction region.

The paper is organized as follows. In Sec. \ref{sec:MM}, we introduce our TDSE modeling of the nonlinear polarization as well as the phenomenological models currently discussed in the literature. For sake of simplicity, we focus on 1D TDSE description of atomic hydrogen. Then, we confront the phenomenological models (HOKE and Standard model) with rigorous TDSE results in Sec. \ref{sec:Comp1D}. Finally, we confirm that our findings hold for more realistic 3D TDSE calculations in Sec. \ref{sec:3D}.

\section{Modeling Nonlinear Polarization\label{sec:MM}}

\subsection{1D Quantum Mechanical Treatment}

Let us start with 1D quantum mechanical calculations. We assume that the single non-relativistic electron of the 1D hydrogen model in an external laser $E(t)$ is described by its wave function $\Psi(x,t)$, governed by the TDSE
\begin{equation}
    i\hbar\partial_t\Psi(x,t)=\left(H_0+H_{int}\right)\Psi(x,t).
    \label{eq:TDSE}
\end{equation} 
Here $H_0=-\hbar^2\partial_x^2/(2m_e)+V(x)$ with electron mass $m_e$, and the interaction with the external laser field in velocity gauge is formulated in the dipole approximation, i.e. $H_{int}=i\hbar q_eA(t)\partial_x/m_e$ involving the electron charge $q_e$ and the magnetic potential $A(t)=-\int_{-\infty}^t E(t')dt'$. We use a soft-core potential $V(x)=-q_e^2/(4\pi\epsilon_0\sqrt{x^2+2a_0^2})$ adjusted to match the ionization energy of hydrogen, where $a_0$ is the Bohr radius.

The macroscopic polarization is then given by $P(t)=\rho_{at}q_e\langle\Psi(t)\lvert \hat{x} \rvert\Psi(t)\rangle$, where $\rho_{at}$ is the atomic density at ambient pressure.

As proposed in \cite{Nurhuda:NJP:10:053006}, we distinguish between bound $\Psi_B$ and continuum $\Psi_C$ electrons by projecting the electronic wavefunction $\Psi$  on bound eigenfunctions of the field free Hamiltonian $H_0\vert\psi_n\rangle=E_n\vert\psi_n\rangle$, such that
\begin{equation}
    \lvert\Psi_B\rangle=\sum_n\lvert\psi_n\rangle\langle\psi_n\rvert\Psi\rangle, \qquad \lvert\Psi_C\rangle=\lvert\Psi\rangle-\lvert\Psi_B\rangle,
    \label{eq:PsiBoundCont}
\end{equation}   
where we used states up to principal quantum number $n=52$. Then, the macroscopic polarization $P(t)$ can be decomposed into  bound, bound-continuum and continuum part, $P(t) = P_{BB}(t)+
2\Re[P_{BC}(t)] + P_{CC}(t)$, according to Eq. \eqref{eq:PsiBoundCont}. For example, 
\begin{equation}
P_{CC}(t)=\rho_{at}q_e\langle \Psi_C(t)\lvert\hat{x}\rvert\Psi_C(t)\rangle\label{eq:PCC}.
\end{equation}

To observe the dependence of $P(t)$ on the laser pulse intensity, we numerically integrate Eq. \eqref{eq:TDSE} for Gaussian pulses 
\begin{equation}
    E(t)=E_0 e^{-(t/\tau)^2}\cos\left(\omega_0t\right) \label{eqn:gaussian}
\end{equation}
with central frequency $\omega_0=2\pi c/\lambda_0$ ($\lambda_0=800$~nm, i.e. cycle period of $2.67$~fs), and peak intensities $I\sim\lvert E_0\rvert^2$ varying from $10^{-3}$~TW/cm$^2$ to $70$~TW/cm$^2$ for pulse durations $\tau=10$~fs, $20$~fs and $100$~fs. This intensity range ensures partial ionization by multi-photon and/or tunnel processes as complete ionization for hydrogen appears at $I=120$~TW/cm$^2$ by over-the-barrier ionization~\cite{Bandrauk:IRAMP:2:1}.

In order to access the macroscopic nonlinear polarization $P_{NL}$, we first extract the linear response $P_{\textrm{Lin}}(t)=\epsilon_0\chi^{(1)}E(t)$, where we neglect linear dispersion and absorption, from low intensity calculations $I<1$~TW/cm$^2$ yielding $\chi^{(1)}=6.88\times10^{-4}$. Then, the  nonlinear part of the polarization can be obtained with  $P_{NL}(t)=P(t)-\epsilon_0\chi^{(1)}E(t)$. For the choice of the carrier envelope phase offset in our test pulses [Eq. \eqref{eqn:gaussian}], $\Re[P_{NL}\left(\omega_0\right)]$ is responsible for self-focusing ($\Re[P_{NL}\left(\omega_0\right)]>0$) or self-defocusing ($\Re[P_{NL}\left(\omega_0\right)]<0$) action, i.e., a positive (resp. negative) nonlinear induced refractive index change. $\Im[P_{NL}\left(\omega_0\right)]$ describes nonlinear absorption, which is weak and will be neglected throughout this paper. 

\subsection{Phenomenological Models}

Let us now investigate the possibility to mimic the behavior of the nonlinear polarization from TDSE calculations by means of approximate models. Such phenomenological description of the nonlinear response is extremely important for the efficient numerical modeling of high intense laser pulse propagation. As already mentioned above, we will approach the TDSE results in terms of two controversial models, currently used in the context of femtosecond filamentation. These models involve only a single set of few parameters for all pulse durations and intensities. 
First, the HOKE model 
\begin{align}
    P_{NL}(t)&=\epsilon_0\chi^{(3)}E(t)^3+\epsilon_0\chi^{(5)}E(t)^5+\epsilon_0\chi^{(7)}E(t)^7+\cdots\nonumber\\
             &=\epsilon_0\sum_{j=3,5,\cdots}\chi^{(j)}E(t)^j
    \label{eq:HOKE}
\end{align}
is fully determined by the nonlinear coefficients $\chi^{(j)}$. Second, the Standard model
\begin{equation}
    P_{NL}(t)=\epsilon_0\chi^{(3)}E(t)^3-q_e^2/\left(m_e\omega_0^2\right)\rho (t) E(t),
    \label{eq:standard}
\end{equation}
requires an additional equation to describe the free electron density 
\begin{equation}
\partial_t\rho(t)=w(t)\left(\rho_{at}-\rho(t)\right).
\end{equation}
Here, we will either apply a multi-photon ionization (MPI) rate \cite{Popov:PU:47:855} 
\begin{equation}
w=w_{MPI}(I(t))=\sigma_KI(t)^K
\end{equation}
with photon number $K$ and ionization cross section $\sigma_K$ or a field-dependent tunnel ionization rate \cite{Landau:1977,Yudin:PRA:64:013409}
\begin{equation}
w=w_{tunnel}(E(t))=\alpha/\lvert E(t)\rvert\exp\left(-\beta/\lvert E(t)\rvert\right)\label{eq:tunnelrate}
\end{equation}
involving two parameters $\alpha$, $\beta$.
\begin{figure*}
    \includegraphics[width=\textwidth]{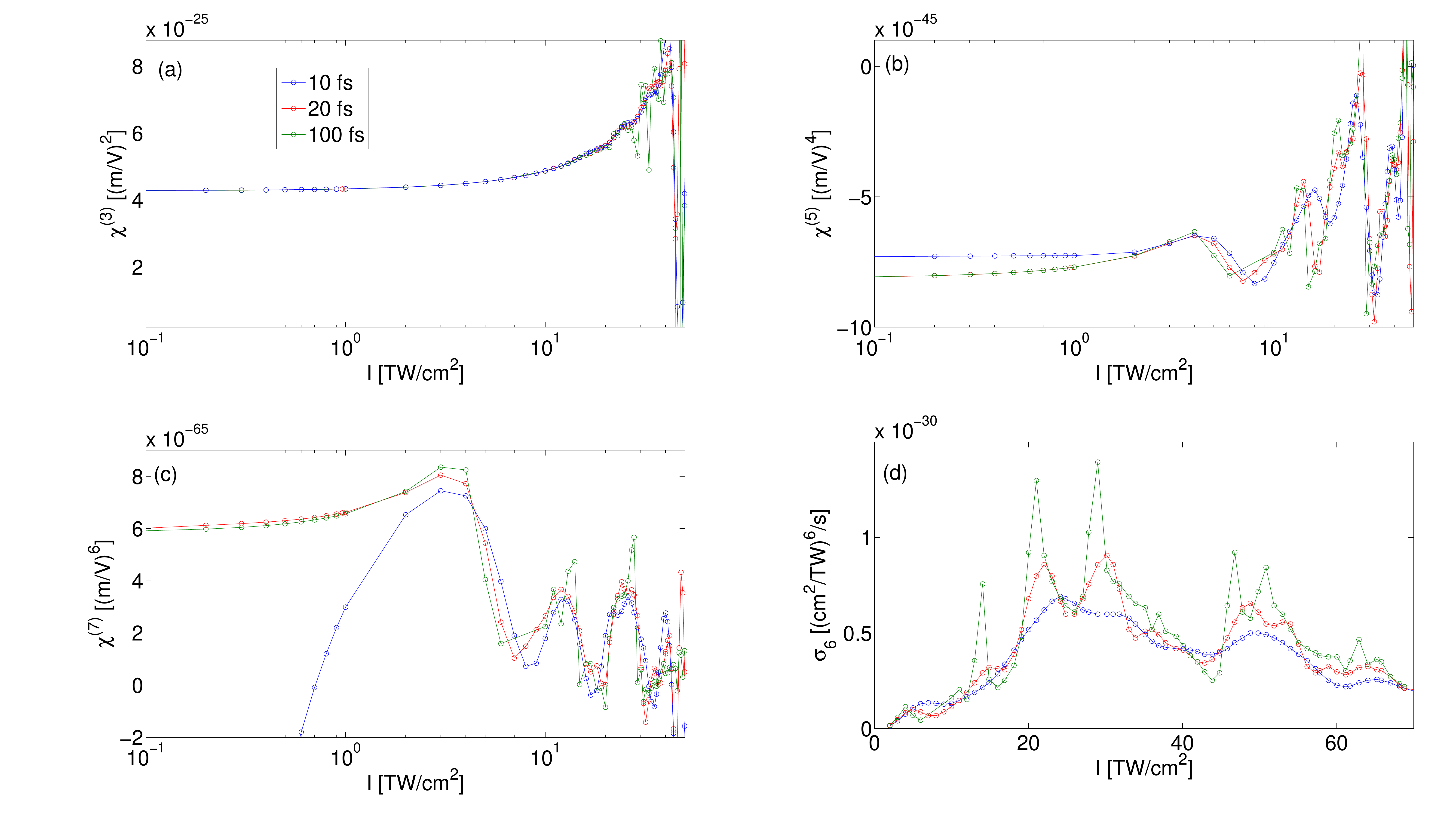}
    \caption{\label{fig:chissigma} Model parameters extracted from 1D TDSE simulations. (a) $\chi^{(3)}$, (b) $\chi^{(5)}$, (c) $\chi^{(7)}$ for $\tau=10$~fs (blue), $\tau=20$~fs (red) and $\tau=100$~fs (green) as function of the peak intensity $I$. (d) The MPI cross section $\sigma_6$ extracted from 1D TDSE simulations [same color coding as in (a)-(c)].}
\end{figure*}

\subsection{Extraction of Phenomenological Model Parameters from 1D TDSE Results\label{sec:extractparam}}

In order to confront the phenomenological models with TDSE results, we first extract the susceptibilities $\chi^{(j)}$ for the HOKE model and additionally the MPI cross section $\sigma_K$ resp. the tunnel ionization parameters $\alpha$, $\beta$ for the two Standard models from the quantum mechanical data. 

For the extraction of the $\chi^{(j)}$ we use the assumption that different orders $P_{NL}^{(j)}(t)=\epsilon_0\chi^{(j)}E(t)^j$ of the nonlinear polarization become important subsequently for increasing field strength.  Additionally, we assume that the characteristic peak at the $j$th harmonic $j\omega_0$ in $\hat{P}^{(j)}_{NL}$ is mainly caused by $\chi^{(j)}$. Then, one can deduce consecutively the $\chi^{(j)}$ using 
\begin{equation}
\hat{P}_{NL}^{(j)}(j\omega_0)=\epsilon_0\chi^{(j)}\sqrt{\pi}\left(\frac{E_0}{2}\right)^{j}\tau\frac{1}{\sqrt{j}}\label{eq:jomega},
\end{equation}
valid for the Gaussian pulses of Eq. \eqref{eqn:gaussian}. Alternatively, we can evaluate $\hat{P}_{NL}(\omega_0,I)$ at different peak intensities $I$ and infer the values of $\chi^{(j)}$ by solving a system of linear equations
\begin{align}
\hat{P}_{NL}(\omega_0,I_1)&=\epsilon_0FT\{\chi^{(3)}E_1^3(t)+\chi^{(5)}E_1^5(t)+\cdots\}\rvert_{\omega_0},\nonumber\\
\hat{P}_{NL}(\omega_0,I_2)&=\epsilon_0FT\{\chi^{(3)}E_2^3(t)+\chi^{(5)}E_2^5(t)+\cdots\}\rvert_{\omega_0},\nonumber\\
\cdots&=\cdots\label{eq:SoEX},
\end{align}
where $FT\{\}$ denotes the Fourier transform. Note, that the former approach evaluates the susceptibilities at different harmonic frequencies $j\omega_0$, whereas the latter one gives access to all $\chi^{(j)}$ directly at $\omega_0$. For purely instantaneous HOKE terms of Eq. \eqref{eq:HOKE}, both approaches should yield the same $\chi^{(j)}$.

Let us first evaluate $\chi^{(j)}$ at the harmonic frequencies $j\omega_0$. Results are presented in Figs. \ref{fig:chissigma} (a)-(c), where $\chi^{(3)}$, $\chi^{(5)}$ and $\chi^{(7)}$ are plotted for different pulse durations $\tau$ vs. peak intensity. We can observe, how difficult it is to extract meaningful coefficients $\chi^{(j)}$ being independent of intensity and pulse duration in the range where $P_{NL}$ becomes relevant, i.e. $I>1$~TW/cm$^2$. The fact that it is not possible to describe the nonlinear polarization in terms of a Taylor expansion at intensities above 10 TW/cm$^2$ already indicates significant changes in the electronic configuration of the atomic system. This involves, e.g., the enhanced population of Stark shifted Rydberg states, staying close to the ionization threshold due to a comparable increase of the ponderomotive potential \cite{Agostini:PRL:63:2208}. Moreover, the observed dependency of the higher order $\chi^{(j)}$ on the pulse duration reflects strong nonlinear dispersion, i.e., a noninstantaneous character of the nonlinear response at higher intensities.
Due to this nonlinear dispersion, and because we are searching for a description of $P_{NL}$ at $\omega_0$, we resort to the latter approach of Eqs. \eqref{eq:SoEX} in the following. By doing so, we can only extract a minimum set of consistent parameters, namely
\begin{align}
\chi^{(3)}&=2.3\times10^{-25} \left(\textrm{m/V}\right)^2\label{eq:HOKE3},\\
\chi^{(5)}&=-2.5\times10^{-45}\left(\textrm{m/V}\right)^4\label{eq:HOKE5},
\end{align}
all higher order $\chi^{(j)}$ being discarded, since all of them strongly fluctuate along the intensity range. We will further refer to this simple model as HOKE\_35.  Note that we will later introduce another set of parameters, HOKE\_FIT, which includes nonlinear coefficients up to $\chi^{(9)}$, chosen to obtain a more accurate description of the nonlinear saturation.
 
For the Standard model with MPI rate, in addition to $\chi^{(3)}$, the photon number $K$ and the cross section $\sigma_K$ has to be determined from the TDSE simulation data. For pulses with wavelength $\lambda_0=800$~nm and the ionization potential of hydrogen of $U_H=13.6$~eV one obtains a minimum photon number of $K=\mod\left(U_H/\hbar\omega_0+1\right)=9$. However, we find the least dependence of $\sigma_K$ on the peak intensity when employing a reduced photon number of $K=6$. This is plausible, since for the here employed intensities the Keldysh parameter $\gamma$ is in the order of one, indicating a transient regime between MPI  ($\gamma\gg 1$) and tunnel ionization ($\gamma\ll 1$), which leads to a reduced exponent in the intensity dependence of the ionization probability. In addition, present atomic resonances connecting the ground and first excited state are of similar order (five and seven photon resonance in 1D and 3D, respectively) \cite{Bian:PRA:83:041403,Bian:PRL:105:093903}. Because, we are only interested in the best (phenomenological) description of TDSE results, we will employ the MPI rate with $K=6$ in the following. To calculate $\sigma_6$, we match the free electron density at the end of the pulse ($t=t_{end}$) obtained with the MPI rate with the TDSE one:
\begin{equation}
\rho(t\rightarrow t_{end})\stackrel{!}{=}\int \textrm{d}x \lvert\Psi_C(x,t_{end})\rvert^2,
\end{equation}
where $\Psi_C(x,t)$ is the continuum part of the wavefunction. Interestingly, the results for all three pulse durations agree, up to strong modulations of $\sigma_6$ at increasing intensity [Fig. \ref{fig:chissigma} (d)]. We attribute these modulations to resonant excitation of high lying Rydberg states, where the energy of an integer number of photons matches the energy gap between the ground and a (Stark shifted) Rydberg state \cite{Volkova:JETPL:94:519}. We thus select an average value for the Standard model,
\begin{equation}
\sigma_6=8.6\times10^{-31}~(\textrm{cm}^2/\textrm{TW})^{6}/\textrm{s}\label{eq:sigK},
\end{equation}
matching $\sigma_6$ at $I=29$~TW/cm$^2$.  Our choice offers a reasonably good fit of the nonlinear polarization close to the clamping intensity, where $P_{NL}$ changes sign. Below this intensity charge contributions are negligible, thus deviating $\sigma_6$ do not matter. For higher intensities we will overestimate the charge, however, at least in the context of femtosecond filamentation, this regime is expected to be barely relevant as well. In any case, complete ionization of the hydrogen atom is expected at $I>1.2\times10^{14}$~W/cm$^2$ by
over-the-barrier ionization~\cite{Bandrauk:IRAMP:2:1}.

To extract the tunnel ionization parameters we calculate the free electron density at the end of the pulse according to the tunnel ionization rate for all peak intensity values employed in the TDSE simulations and compare $\rho(t\rightarrow t_{end})$ to the TDSE data for $\tau=20$~fs. Then, $\alpha$ and $\beta$ are optimized such that the sum of the squared differences from each comparison is minimized. Performing this least squares fit, we extract 
\begin{align}
    \alpha &=8.9\times10^{26} \textrm{V/ms},\label{eq:alpha}\\
    \beta &=1.6\times10^{11} \textrm{V/m},\label{eq:beta}
\end{align}
for the parameters entering Eq. \eqref{eq:tunnelrate}. These values are used for all pulse durations in the following. 
\begin{figure*}
    \includegraphics[width=\textwidth]{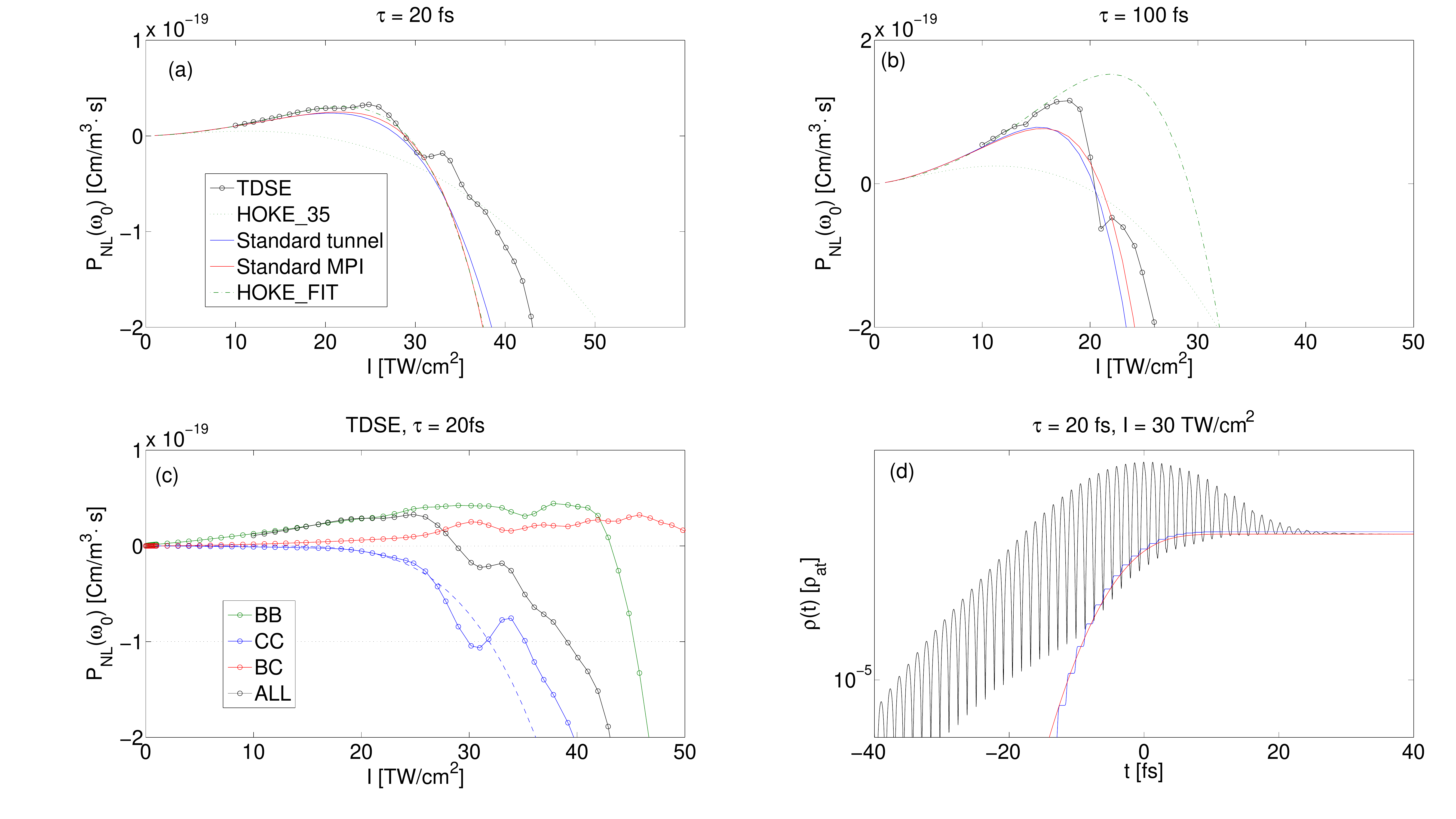}
    \caption{\label{fig:1DResults_FD} 1D simulation results. For (a) $\tau=20$~fs and (b) $\tau=100$~fs the nonlinear polarization at $\omega_0$ is shown for TDSE calculations (solid black lines with circles) and compared to the Standard models (solid red lines $\equiv$ MPI rate, solid blue $\equiv$ tunnel rate), the HOKE\_35 model (dotted green lines) containing a $\chi^{(3)}$ and $\chi^{(5)}$ contribution only and the fitted HOKE\_FIT model (dash-dotted green lines). (c) distinguishes bound (green), continuum (blue) and bound-continuum (red) electronic contributions to the nonlinear polarization (black) for $\tau=20$~fs (see text for details). Additionally, the continuum contribution from the Standard model with tunnel ionization rate is shown (dashed blue line). In (d) the free electron density according to TDSE simulations (black) is compared to the one obtained from the MPI rate (red) and tunnel rate (blue) as used in the Standard models.}
\end{figure*}

\section{Confrontation of phenomenological models with 1D TDSE results\label{sec:Comp1D}}

\subsection{Behavior in Fourier Domain}

Having all necessary parameters at hand, let us now confront the HOKE and plasma-based (Standard) description of the nonlinear polarization with TDSE results. In Figs. \ref{fig:1DResults_FD} (a) and (b), we compare the nonlinear polarization $\hat{P}_{NL}(\omega_0)$ at center frequency of the laser obtained from TDSE simulations with the approximate models for $\tau=20$~fs and $\tau=100$~fs, respectively. Shorter pulses ($\tau=10$~fs) yield similar behavior (not shown). For the TDSE results with $\tau=20$~fs (solid black line) $\hat{P}_{NL}(\omega_0)$ is positive and increases (focusing action) with peak intensity up to $I\simeq25$~TW/cm$^2$. Then, saturation and final change of sign occur at $I_{\rm sat}\simeq 30$~TW/cm$^2$, signaling a qualitative change from focusing to defocusing behavior. This generic behavior also holds for $\tau=100$~fs with decreased [$I_{\rm sat}\simeq20$~TW/cm$^2$, Fig. \ref{fig:1DResults_FD} (b)] and $\tau=10$~fs with increased [$I_{\rm sat}\simeq40$~TW/cm$^2$, not shown] saturation intensity values. Moreover, we observe oscillatory structures of $\hat{P}_{NL}(\omega_0)$ at $I\simeq35$~TW/cm$^2$ ($\tau=20$~fs) and $I\simeq22$~TW/cm$^2$ ($\tau=100$~fs), which can be attributed to decreased ionization probabilities due to channel closure \cite{Kopold:JPB:35:217}. The HOKE\_35 model, defined by extracted $\chi^{(3)}$ and $\chi^{(5)}$ coefficients [see Eqs. \eqref{eq:HOKE3}, \eqref{eq:HOKE5}] (green dotted line) yields significantly lower intensity thresholds for saturation and sign inversion of $\hat{P}_{NL}(\omega_0)$ than observed in the TDSE calculations.
Because this discrepancy renders the HOKE\_35 unsuitable for practical applications, we suggest a different (purely phenomenological) set of parameters HOKE\_FIT
\begin{align}
 \chi^{(3)}&=2.3\times10^{-25}\left(\textrm{m/V}\right)^2,\\
\chi^{(5)}&=-2.0\times10^{-45}\left(\textrm{m/V}\right)^4,\\
\chi^{(7)}&=5.3\times10^{-65}\left(\textrm{m/V}\right)^6,\\
\chi^{(9)}&=-2.9\times10^{-85}\left(\textrm{m/V}\right)^8,
\end{align}
entering Eq. \eqref{eq:HOKE} which satisfactorily reproduces the TDSE data for $\tau=20$~fs. However, changing the pulse duration destroys this agreement [see results for HOKE\_FIT, green dash-dotted line in Fig. \ref{fig:1DResults_FD} (a), (b)]. In contrast, comparison to the Standard models with either ionization rate (red line $\equiv$ MPI rate, blue $\equiv$ tunnel rate) shows reasonable agreement for the physically extracted model parameters [see Eqs. \eqref{eq:HOKE3} and \eqref{eq:sigK}-\eqref{eq:beta}] for all pulse durations. 

An analysis of the physical mechanism responsible for the observed nonlinear refractive index saturation in the TDSE simulations is detailed in Fig. \ref{fig:1DResults_FD} (c). Here, the resulting contributions to the nonlinear polarization from the bound ('BB' green line), continuum ('CC' blue line) and bound-continuum ('BC' red line) electronic wave function [see 
Eq. \eqref{eq:PCC}] are shown for $\tau=20$~fs. The bound (BB) and bound-continuum (BC) contributions to $P_{NL}(\omega_0)$ are both positively valued (focusing action) up to 45 TW/cm$^2$, well beyond the threshold intensity where the overall nonlinear polarization (solid black line) changes sign. In contrast, the continuum contribution (CC) is close to zero for small intensities $I<20$~TW/cm$^2$, because no ionization occurs, and yields a negative (defocusing) contribution when free charge generation kicks in. The value of this negative contribution (CC) yields significantly lower intensity thresholds than both the positive contributions of BB and BC. Thus, the observed sign flip in the nonlinear polarization, i.e., the saturation of the nonlinear induced refractive index, can be clearly attributed to the contribution of free electrons {\it according to} Eq. \eqref{eq:PCC}. In this context it is worth mentioning that, strictly speaking, only the overall polarization is gauge invariant. The projected contributions (BB, BC, CC) may differ for a different gauge in the TDSE equation. However, in particular the CC contribution preserves its qualitative focusing properties using different gauges, as shown in \cite{Bejot:arxiv:2012}.
 
To quantify the impact of the free electrons, ionization is considered in Fig. \ref{fig:1DResults_FD} (d). The free electron density obtained from TDSE simulations via $\rho(t)=\rho_{at}\int\textrm{d}x\lvert\Psi_C(x,t)\rvert^2$ [shown for $I=30$~TW/cm$^2$ and $\tau=20$~fs; solid black line] develops strong oscillations at times where the pulse interacts with the atom. This is due to the fact that our definition using Eq. \eqref{eq:PsiBoundCont} for the free electron density is only applicable for the field free situation after the driving pulse has passed. In the presence of the laser field the eigenstates of $H_0$ do no longer represent the bound states of the laser dressed atom. Therefore, spurious oscillations occur due to contributions from electrons in high lying bound (Rydberg/Kramers-Henneberger) states close to the bulge of the bended atomic potential {\it and} from truly free electrons. However, in the Standard models we apply ionization rates that describe the truly ionized fraction {\it after} the laser pulse only, and the corresponding charge contributions do not feature these oscillations [red line $\equiv$ MPI rate, blue line $\equiv$ tunnel rate in Fig. \ref{fig:1DResults_FD} (d)]. Importantly, the truly ionized fraction according to the Standard models [dashed blue line for tunnel rate in \ref{fig:1DResults_FD} (c)] reproduces the CC contributions from TDSE simulations for all intensities and pulse durations. Minor deviations for $I\gtrsim30$~TW/cm$^2$ are due to previously discussed resonant occupation of high lying Rydberg states \cite{Volkova:JETPL:94:519}, whose description is beyond the scope of simple MPI/tunnel ionization models.  

\begin{figure*}
     \includegraphics[width=\textwidth]{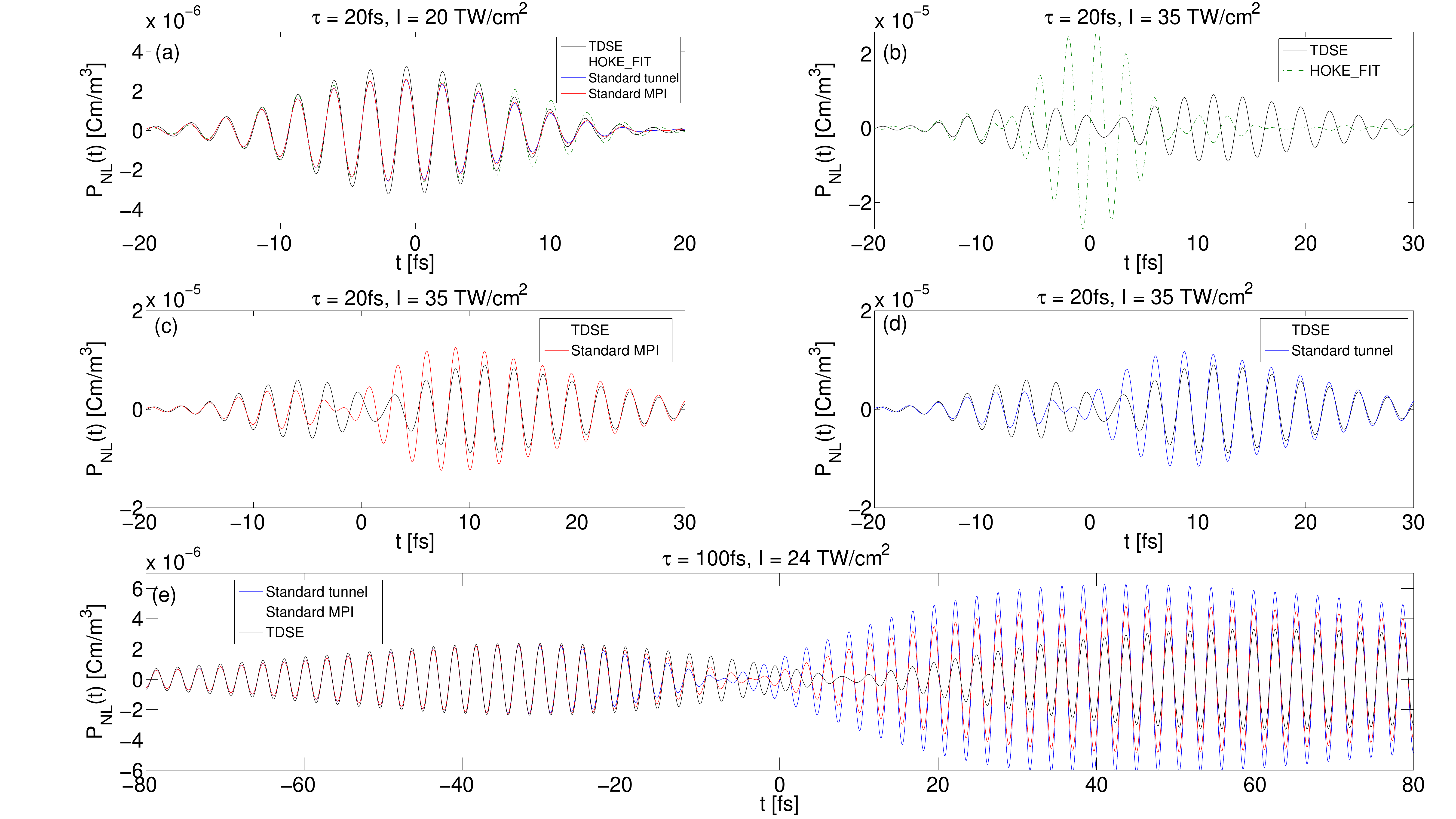}
     \caption{\label{fig:PNLTimeDomain} 1D nonlinear polarization in time domain. (a) Results from TDSE and approximate models for $\tau=20$~fs at $I=20$~TW/cm$^2$, where free electron contributions are negligible and the response is governed by the instantaneous contribution from bound electrons. (b)-(e) Temporal shapes, where delayed charge contributions become important: (b) TDSE and HOKE\_FIT, (c) TDSE and Standard model with MPI rate and (d) TDSE and Standard model with tunnel rate for $\tau=20$~fs and $I=35$~TW/cm$^2$; (e) TDSE and MPI/tunnel Standard models for $\tau=100$~fs and $I=24$~TW/cm$^2$. All temporal field shapes were filtered in Fourier domain using a hyperbolic tangent mask around $0.8\omega_0<\omega<1.2\omega_0$. Color coding is the same as in Fig. \ref{fig:1DResults_FD}.}
\end{figure*}

\subsection{Behavior in Temporal Domain}

Let us continue our comparison of TDSE results for $P_{NL}$ with approximate models in time domain. In Fig. \ref{fig:PNLTimeDomain} (a) the time evolution of $P_{NL}$ from TDSE simulations and the approximate HOKE\_FIT and Standard models are shown for $\tau=20$~fs and $I=20$~TW/cm$^2$. For that intensity, charge contributions are small and all three models reproduce accurately the response of the bound electrons, instantaneously following the electric field. For $I=35$~TW/cm$^2$ in (b) charge contributions become important, resulting in a delayed response of opposite sign kicking in at $t\sim5$~fs, according to TDSE simulations (black solid line). The HOKE\_FIT model (dash-dotted green line), although yielding a comparable value of $\hat{P}_{NL}(\omega_0)$, cannot describe that delayed contribution due to its instantaneous nature and produces much higher polarization amplitudes than observed in the TDSE simulations. In contrast, both Standard models in (c), (d) qualitatively capture this delayed response and they develop only limited differences in amplitude. A similar situation is encountered in (e) for $\tau=100$~fs and $I=24$~TW/cm$^2$, where also the limits of the Standard models are illustrated more clearly: The qualitative temporal character is described correctly over long evolution times (oscillation in phase with TDSE results). However, due to a slightly overestimated charge contribution, the Standard models yield larger amplitudes for the delayed response coming along with an inaccurate onset of the latter.      

\section{Comparing to 3D TDSE Calculations\label{sec:3D}}

Finally, let us confirm the results obtained from the 1D models by performing the same analysis in 3D. Here, we numerically solve 
\begin{equation}
    i\hbar\partial_t\Psi(\vec{r},t)=\left(H_0+H_{int}\right)\Psi(\vec{r},t)\label{eq:TDSE3D}
\end{equation} 
with $H_0=-\nabla^2/(2m_e)+V(r)$ including $V(r)=-q_e^2/4\pi\epsilon_0r$ with $r=\lvert\vec{r}\rvert$. The interaction with the $x$-linearly polarized external laser field in length gauge is computed in dipole approximation $H_{int}=-q_eE(t)x$ \cite{Cormier:JPB:30:77}. We followed the same procedure to extract the parameters of the approximate models as in the 1D case  (see Sec. \ref{sec:extractparam}).

\begin{figure*}
    \includegraphics[width=\textwidth]{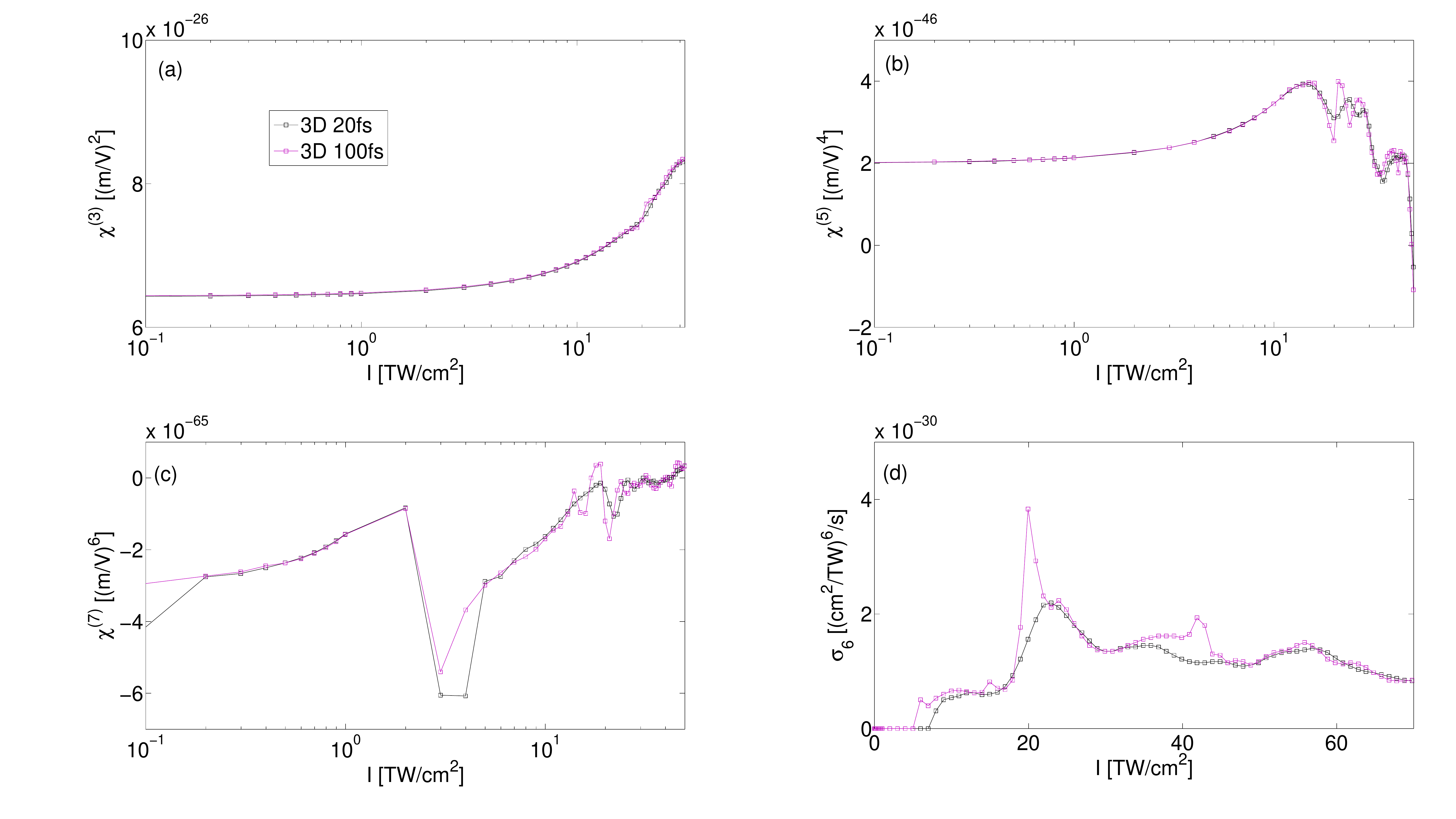}
    \caption{\label{fig:chissigma3D} Model parameters extracted from 3D TDSE simulations. (a) $\chi^{(3)}$, (b) $\chi^{(5)}$, (c) $\chi^{(7)}$ for $\tau=20$~fs (black), $\tau=100$~fs (magenta) as function of the peak intensity $I$. (d) The MPI cross section $\sigma_6$ extracted from 3D TDSE simulations [same color coding as in (a)].}
\end{figure*}

From low intensity calculations we obtain a linear susceptibility of $\chi^{(1)}=2.3\times10^{-4}$. The nonlinear susceptibilities $\chi^{(3)},\chi^{(5)},\cdots$ evaluated at harmonic frequencies [see Eq. \eqref{eq:jomega}] are presented in Figs. \ref{fig:chissigma3D} (a)-(c) for pulse durations of $\tau=20$~fs (solid black line) and $\tau=100$~fs (solid magenta line). The deviations for different $\tau$ are similar to those observed in the 1D case, indicating similar nonlinear dispersion. Because we again meet difficulties in extracting a consistent set of $\chi^{(j)}$s from Eqs. \eqref{eq:SoEX}, we omit a repeated comparison to the HOKE\_35 model. Thus we resort to comparisons involving the HOKE\_FIT model, defined through 
\begin{align}
 \chi^{(3)}&=0.65\times10^{-25}\left(\textrm{m/V}\right)^2,\\
\chi^{(5)}&=0.13\times10^{-45}\left(\textrm{m/V}\right)^4,\\
\chi^{(7)}&=-1.0\times10^{-66}\left(\textrm{m/V}\right)^6,\\
\chi^{(9)}&=-1.0\times10^{-87}\left(\textrm{m/V}\right)^8,
\end{align}
and the Standard models employing
\begin{equation}
\chi^{(3)}=0.65\times10^{-25}\left(\textrm{m/V}\right)^2 
\end{equation}
together with ($K=6$)
\begin{equation}
\sigma_6=1.13\times10^{-30}(\textrm{cm}^2/\textrm{TW})^{6}/\textrm{s}
\end{equation}
for the MPI rate or
\begin{align}
    \alpha &=2.0\times10^{26} \textrm{V/ms},\\
    \beta &=1.9\times10^{11} \textrm{V/m}
\end{align}
for the tunnel ionization rate. Naturally, the values of the model parameters in 3D differ from the ones obtained from 1D TDSE data. Nevertheless, our 3D calculations yield the same order of magnitude for the $\chi^{(j)}$'s as observed in 1D. It is interesting that $\chi^{(5)}$ and $\chi^{(7)}$  exhibit opposite signs compared to the 1D case.  

\begin{figure*}
     \includegraphics[width=\textwidth]{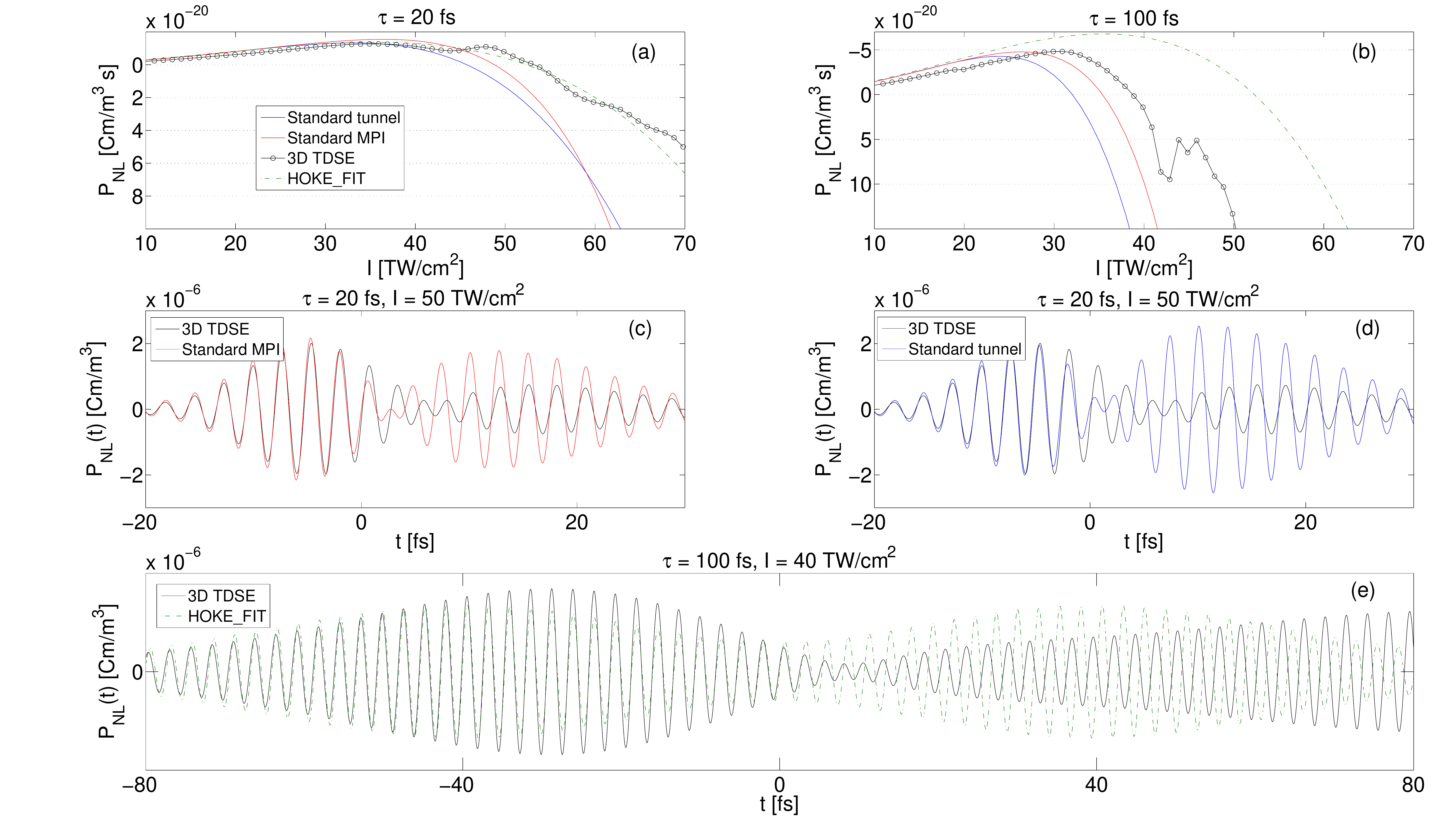}
     \caption{\label{fig:3D} 3D simulation results. $P_{NL}(\omega_0)$ for (a) $\tau=20$~fs and (b) $\tau=100$~fs. $P_{NL}(t)$ from 3D simulations and Standard model with (c) MPI rate and (d) tunnel rate for $\tau=20$~fs, $I=50$~TW/cm$^2$. (e) 3D simulations and HOKE\_FIT for $\tau=100$~fs, $I=40$~TW/cm$^2$. The pulse intensities are chosen to evidence the delayed nature of charge contributions.}
\end{figure*}

Let us continue with the comparison of 3D TDSE data with HOKE\_FIT and Standard model results. In Fig. \ref{fig:3D} (a) for $\tau=20$~fs and (b) for $\tau=100$~fs  the nonlinear polarization $\hat{P}_{NL}(\omega_0)$ inferred from 3D simulations (solid black line) is compared to the HOKE\_FIT model (dash-dotted green line) and to the Standard models with a MPI rate (solid red line) or a tunnel ionization rate (solid blue line). As in the 1D case, the HOKE\_FIT model exhibits strong deviations for different pulse durations $\tau$. In contrast, the Standard models with physically extracted parameters reasonably approximate simulation results for both pulse durations $\tau$. However, for intensities $I\gtrsim50$~TW/cm$^2$ for the $\tau=20$~fs case and  $I\gtrsim30$~TW/cm$^2$ for $\tau=100$~fs the overestimation of charge contributions seems more pronounced than in 1D. In Fig. \ref{fig:3D} (c) and (d) this is evident when comparing the temporal evolution of $P_{NL}(t)$ from 3D TDSE simulations (solid black line) to Standard models with (c) MPI (red solid line) and  (d) tunnel ionization rate (blue solid line). Here, we show data for $\tau=20$~fs and $I=50$~TW/cm$^2$, which ensures a considerable charge contribution. Both Standard models qualitatively reproduce simulation results by capturing the instantaneous and delayed property. However, deviations in the amplitude of the delayed contribution are larger than in similar 1D situations [Fig. \ref{fig:PNLTimeDomain} (c), (d)]. The HOKE\_FIT model (dash-dotted green line) in Fig. \ref{fig:3D} (e) for $\tau=100$~fs and $I=40$~TW/cm$^2$ not even qualitatively captures the delayed temporal property as it gives out of phase oscillations with an opposite sign for the field amplitude for $t\gtrsim10$~fs. 

\section{Conclusion}
 
In conclusion, numerical solution of the TDSE for atomic hydrogen in 1D and 3D reveals the saturation and change of sign of the nonlinear polarization at central laser frequency for rising intensity of the driving pulse. As in \cite{Bejot:arxiv:2012}, we identify a time retarded contribution responsible for the transition from the focusing to defocusing regime. This time retarded contribution cannot be modeled by means of an instantaneous HOKE model, as we neither can  physically extract consistent HOKE terms nor (even qualitatively) reproduce temporal field structures from simulations with a phenomenologically fitted HOKE\_FIT model. On the other hand, applying Standard models with either a MPI or tunnel ionization rate, clearly attributes the delayed response to the generation of free electrons. With both standard models we achieve reasonable agreement with TDSE simulations for the nonlinear polarization at central laser frequency as well as for the temporal field structures. We propose a unique set of parameters, valid for all applied pulse durations and intensities, to model the nonlinear polarization. This simple approach opens the possibility to efficiently simulate laser pulse propagation with reasonable accuracy. However, due to their inherently approximate nature, possibly important details observed in TDSE simulations, e.g. the non-monotonous behavior of the ionization probability due to channel closure and resonant excitation, are not included in the Standard models. Thus, in order to accurately describe laser pulse propagation, a coupling of TDSE calculations to propagation equations as in \cite{Lorin:PhysD:241:1059} seems mandatory.  

We gratefully thank M. Ivanov, M. Kolesik, A.M. Popov and E. Cormier for fruitful and inspiring discussions on the bound electrons in the continuum.

\bibliography{./references}
\end{document}